\documentclass[12pt]{article}
\usepackage[a4paper,margin=2.5cm]{geometry}
\usepackage{setspace}
\usepackage{url}
\usepackage{hyperref}
\usepackage{graphicx}
\usepackage[numbers]{natbib}
\usepackage{titlesec}
\usepackage{authblk}
\usepackage{float}
\usepackage{graphicx}
\usepackage{caption}

\title{\textbf{Embedding Explainable AI in NHS Clinical Safety: The Explainability-Enabled Clinical Safety Framework(ECSF)}}

\author{Robert Gigiu, RN}
\affil{Independent Researcher, United Kingdom\\
\href{https://orcid.org/0009-0007-4093-9230}{ORCID: 0009-0007-4093-9230}}

\begin{document}

\maketitle

\begin{abstract}

\textbf{Background:} 
Artificial intelligence (AI) is increasingly embedded in NHS workflows. Its probabilistic and adaptive behaviour conflicts with the deterministic assumptions underpinning existing clinical-safety standards. Current NHS standards DCB0129 and DCB0160 provide strong governance for deterministic software but do not define how AI-specific transparency, interpretability, or model drift should be evidenced within Safety Cases, Hazard Logs, or post-market monitoring.

\textbf{Objective:} 
To propose an Explainability-Enabled Clinical Safety Framework (ECSF) that integrates explainability into the DCB0129 and DCB0160 lifecycle, enabling Clinical Safety Officers to use interpretability outputs as structured safety evidence without creating new artefacts or altering compliance pathways.

\textbf{Methods:} 
A cross-regulatory synthesis mapped DCB0129 and DCB0160 clauses to principles from Good Machine Learning Practice (GMLP), the NHS AI Assurance and T.E.S.T. frameworks, and the EU AI Act. A focused review identified explainable-AI techniques capable of generating auditable evidence across model types. These were synthesised into a conceptual matrix linking regulatory clause, principle, ECSF checkpoint, and acceptable explainability outputs.

\textbf{Results:} 
ECSF introduces five explainability checkpoints aligned with the DCB lifecycle: global transparency for hazard identification, case-level interpretability for verification, clinician-usability evidence for evaluation, traceable decision pathways for risk control, and longitudinal interpretability monitoring for post-market surveillance. Techniques such as SHAP, LIME, Integrated Gradients, saliency mapping, and attention visualisation are mapped to corresponding DCB artefacts. A worked example using a simulated sepsis early-warning system illustrates how explainability evidence can populate safety documentation with verifiable audit trails.

\textbf{Conclusions:} 
ECSF reframes explainability as a core element of clinical-safety assurance. Embedding interpretable evidence within established NHS documentation bridges deterministic risk governance with the probabilistic behaviour of AI and supports alignment with GMLP, the EU AI Act, and NHS AI Assurance principles.

\textbf{Keywords:} 
Explainable artificial intelligence; clinical safety; DCB0129; DCB0160; Safety Case; Hazard Log; post-market surveillance; NHS AI Assurance; GMLP; EU AI Act.

\end{abstract}

\section{Introduction}

Artificial intelligence (AI) has become integral to modern healthcare, supporting diagnostic imaging, triage, and documentation. Its ability to recognise complex data patterns has accelerated innovation but has also exposed limitations in traditional safety assurance. Unlike deterministic medical software, AI systems exhibit non-deterministic and probabilistic behaviour, meaning that identical inputs can yield variable outputs depending on model parameters, training data, or stochastic inference processes \cite{Summers2021,Atil2025}. This inherent variability challenges conventional clinical-safety methodologies, which assume stable and repeatable system logic.

The NHS Digital clinical-safety standards, DCB0129 and DCB0160, provide a structured and risk-based framework for assuring health IT systems \cite{NHS2013a,NHS2013b}. These standards were developed for deterministic software such as electronic health records and decision-support tools, where predictable logic can be verified through testing and validation. As AI technologies increasingly influence diagnosis and treatment, the limitations of deterministic assurance models have become evident. Current standards do not explicitly address algorithmic transparency, data drift, or model interpretability, which are key factors for assessing AI safety \cite{Chustecki2024,Bajwa2021}.

To address these gaps, recent national and international initiatives have advanced complementary principles for AI assurance. The NHS AI Assurance and T.E.S.T. frameworks extend governance to include fairness, accountability, transparency, and post-deployment monitoring \cite{NHSAIAssurance2023,NHSTEST2023}. The  Good Machine Learning Practice (GMLP) guidance defines expectations for data quality, transparency, and human oversight \cite{MHRA2021}, while the EU Artificial Intelligence Act establishes legally binding requirements for high-risk AI systems, including medical devices, with implementation stages beginning in 2025 \cite{EUAIACT2024}. Although these frameworks are conceptually aligned, they remain fragmented and do not yet provide a consistent mechanism for embedding transparency and traceability within formal clinical-risk documentation such as the Safety Case or Hazard Log required under DCB0129 and DCB0160.

Explainability, defined as the ability to make model reasoning observable and interpretable, has emerged as a cornerstone of trustworthy AI. In healthcare, explainability extends beyond transparency by enabling clinical intelligibility, allowing clinicians and regulators to understand not only what an AI system predicts but also why it produces those predictions. Research indicates that explainability enhances clinician confidence, supports validation, and aligns with ethical principles such as autonomy and accountability \cite{Holzinger2019a,Holzinger2019b,Tonekaboni2019,Amann2020}. However, its adoption within formal NHS assurance processes remains limited and largely conceptual. Explainability techniques such as SHAP, LIME, and attention visualisation can make model reasoning auditable, converting probabilistic outputs into interpretable evidence that supports risk identification, control, and validation within the DCB0129 and DCB0160 lifecycle.

This paper introduces the Explainability-Enabled Clinical Safety Framework (ECSF), a model that embeds explainability as a routine assurance mechanism within existing NHS clinical-safety processes. The ECSF reframes explainability from a technical supplement into a core safety function, strengthening traceability, validation, and post-market governance \cite{Hildt2025,Linardatos2021}. It provides a practical foundation for aligning deterministic safety standards with the probabilistic nature of AI systems while maintaining both regulatory and ethical integrity \cite{Bajwa2021,Amann2020}.
\section{Methods}

\subsection{Framework Development Approach}

This study adopted a conceptual modeling approach and cross-regulatory synthesis to integrate explainability principles within existing NHS clinical-safety standards. Rather than proposing a new assurance framework, the analysis sought to align established safety, ethical, and regulatory instruments, including DCB0129 and DCB0160, the GMLP principles, the EU AI Act, and the NHS AI Assurance and T.E.S.T. frameworks \cite{NHS2013a,NHS2013b,MHRA2021,EUAIACT2024,NHSAIAssurance2023,NHSTEST2023}. The objective was to identify practical points of convergence where explainability could enhance clinical risk management and governance.

The development of the framework followed three sequential stages. The first stage, regulatory mapping, involved analysing each clause of the DCB0129 and DCB0160 lifecycle in relation to corresponding principles in the GMLP, the EU AI Act, and the NHS AI Assurance frameworks. This mapping process revealed conceptual overlaps such as traceability, validation, and human oversight, as well as areas of divergence including interpretability requirements and post-market transparency obligations. The findings from this exercise provided the structural foundation for the ECSF.

The second stage, explainability technique mapping, consisted of a  review of literature and regulatory guidance to identify explainable artificial intelligence (XAI) methods capable of generating auditable evidence across different model classes. The review focused on commonly adopted techniques that provide a representative overview of the field, rather than an exhaustive catalogue of all available methods. Priority was given to approaches capable of producing interpretable outputs applicable to DCB artefacts. These included SHAP  and Permutation Importance, which provide quantitative feature attribution for traditional machine-learning models and enable both local and global transparency; LIME, which supports local surrogate modelling and approximate global pattern detection through aggregated samples; and Saliency Maps, Grad-CAM, and Integrated Gradients, which offer visual and gradient-based interpretation for deep-learning models. Additional techniques such as Counterfactual and Contrastive Explanations were evaluated for their ability to explore decision boundaries and support usability assessment.

For large-language and generative models, methods including Rationale Tracing, Token-level Attention Visualisation, and Chain-of-Thought Summarisation were examined for their capacity to capture reasoning logic, contextual reliability, and hallucination risk. Outputs from these methods were conceptually mapped to assurance artefacts required under DCB0129 and DCB0160, namely the Clinical Safety Case, the Hazard Log, and the Post-Market Surveillance Plan. Supporting techniques such as Failure Modes and Effects Analysis (FMEA) were incorporated to demonstrate how Clinical Safety Officers could use interpretability findings to strengthen hazard identification, validation, and mitigation documentation \cite{Ribeiro2016,Lundberg2017,Samek2019,Amann2020,Hildt2025,Asgari2025}.

The third stage, framework construction, involved synthesising insights from the regulatory and technical mappings into the framework. The ECSF introduces five explainability checkpoints aligned with the DCB lifecycle: hazard identification, validation and verification, clinical evaluation, risk control, and post-market surveillance. Each checkpoint specifies acceptable forms of explainability evidence, the corresponding DCB artefact, and the assurance purpose such as validation, traceability, or drift monitoring. Collectively, these checkpoints establish explainability as a structured, repeatable, and measurable component of clinical-safety assurance.

\subsection{Analytical Perspectives}

To ensure coherence between regulatory intent, technical evidence, and clinical usability, the analysis was organised around three complementary perspectives that together informed the construction of the ECSF.

The first perspective, the \textit{regulatory analysis}, examined how explainability principles align with external obligations defined by the GMLP guidance, the EU AI Act, and the NHS AI Assurance frameworks \cite{MHRA2021,EUAIACT2024,NHSAIAssurance2023}. Clauses within DCB0129 and DCB0160 were cross-referenced with these instruments to identify where explainability outputs could provide evidence of compliance with transparency, documentation, and human oversight requirements. Article 13 of the EU AI Act, which mandates transparency and provision of information to deployers, overlaps most strongly with GMLP Principles 1, 9, and 10, and to a lesser extent Principle 3. Together, these principles emphasise multidisciplinary expertise, data representativeness, clear user information, and lifecycle monitoring. This alignment ensures that explainability contributes directly to the evidentiary trail required for regulatory assurance, rather than functioning as a separate technical exercise.

The second perspective, the technical analysis, evaluated how interpretability methods can be embedded within existing clinical safety artefacts. Each phase of the DCB0129 and DCB0160 lifecycle was mapped to a specific ECSF checkpoint and its corresponding form of explainability evidence. For example, global explainability outputs such as aggregated SHAP or Permutation Importance analyses can support hazard identification and bias evaluation within the Hazard Log. Local interpretability techniques such as LIME, Integrated Gradients, and saliency maps can substantiate validation and verification evidence within the Clinical Safety Case. Together, these mechanisms demonstrate how explainability can generate traceable and auditable safety evidence within the established DCB assurance workflow, improving both transparency and reproducibility \cite{Ribeiro2016,Lundberg2017,Samek2019,Amann2020,Asgari2025}.

The third perspective, the \textit{clinical analysis}, considered how explainability affects clinician interpretation, trust, and decision confidence. Based on previous research \cite{Tonekaboni2019,Hildt2025}, this component of the analysis assessed interpretive clarity, cognitive load, and usability of different forms of explanation output. Factors such as visual clarity, consistency of terminology, and alignment with clinical reasoning processes were treated as essential conditions for effective communication of safety evidence. Ensuring that explanations are understandable and proportionate to the needs of the end user helps strengthen human oversight while minimising cognitive burden and the risk of automation bias.

Collectively, these three perspectives ensure that the ECSF is regulatorily aligned, technically verifiable, and clinically intelligible. By integrating regulatory compliance, technical rigour, and human-centred design, the framework supports the systematic use of explainability as an operational component of clinical-safety assurance.

\begin{figure}[H]
\centering
\includegraphics[width=\linewidth]{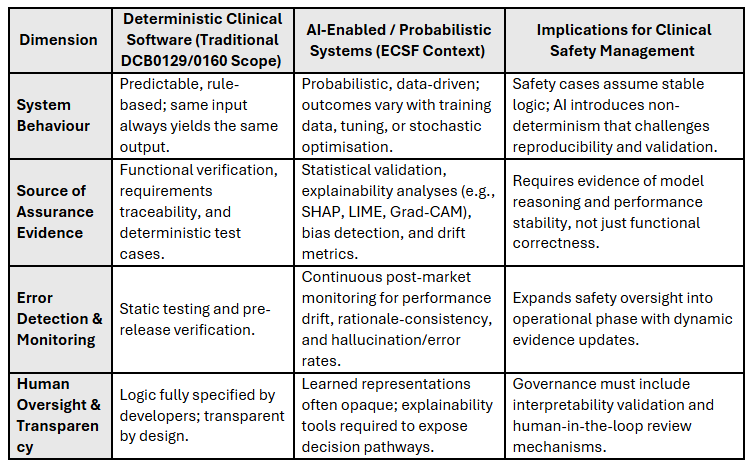}
\caption{Deterministic vs AI-Enabled Clinical Safety Comparison. 
This figure contrasts the assurance logic of traditional deterministic software under DCB0129/0160 with the probabilistic behaviour of AI-enabled systems, 
illustrating why additional explainability and post-market monitoring are required.}
\label{fig:deterministic-vs-ai}
\end{figure}

\subsection{Data Sources}

The conceptual analysis relied on a combination of regulatory, policy, and academic sources to ensure that the ECSF was grounded in both policy compliance and scientific evidence. The regulatory foundation was established through the review of core NHS and governmental standards, including the NHS Digital clinical-safety standards DCB0129 and DCB0160 \cite{NHS2013a,NHS2013b}, the GMLP guiding principles \cite{MHRA2021}, the EU Artificial Intelligence Act \cite{EUAIACT2024}, and the NHS AI Assurance and T.E.S.T. frameworks published between 2023 and 2024 \cite{NHSAIAssurance2023,NHSTEST2023}. These documents provided the structural and procedural requirements against which explainability principles were mapped.

Complementing this regulatory foundation, a review of peer-reviewed literature was undertaken to identify influential academic perspectives on explainable artificial intelligence (XAI) in healthcare. Key works included Holzinger et al.\ \cite{Holzinger2019a}, which outlines the foundational principles of explainability and causability; Tonekaboni et al.\ \cite{Tonekaboni2019}, which explores clinician-centred requirements for interpretable AI; and Amann et al.\ \cite{Amann2020}, which presents a multidisciplinary view of explainability in medical contexts. Additional methodological surveys, including those by Carvalho et al.\ \cite{Carvalho2019}, Linardatos et al.\ \cite{Linardatos2021}, and Hildt \cite{Hildt2025}, provided comprehensive reviews of interpretability techniques and ethical considerations relevant to clinical-safety assurance.

Together, these regulatory and academic sources informed the mapping, synthesis, and validation processes underpinning the ECSF, ensuring that the framework aligns with existing NHS assurance requirements while reflecting the latest advances in explainable-AI research.
\section{Results: The Explainability-Enabled Clinical Safety Framework (ECSF)}

\subsection{Overview}

The ECSF integrates established XAI techniques into the DCB0129 and DCB0160 lifecycle to enhance transparency, traceability, and accountability. The framework aligns the deterministic assurance logic of existing NHS safety standards with the probabilistic behaviour of AI systems and reframes explainability from a research objective into a source of verifiable clinical-safety evidence \cite{Holzinger2019a,Holzinger2019b,Tonekaboni2019}.

As outlined by Arrieta et al.\ \cite{Arrieta2020}, explainability operates at two complementary levels. \textit{Global explainability} concerns understanding how a model behaves overall, including its dominant features, decision boundaries, and potential sources of bias. \textit{Local explainability} focuses on clarifying why a specific output or prediction is generated for an individual case. These two levels of interpretability provide distinct yet complementary forms of assurance within the DCB0129 and DCB0160 lifecycle, supporting both system-level transparency and case-level validation.

DCB0129 and DCB0160 require a documented clinical risk-management process that includes a Hazard Log to record identified hazards, associated controls, and supporting evidence; a Clinical Safety Case to demonstrate that residual risks are acceptable; and post-deployment monitoring and incident management to maintain ongoing safety assurance. \cite{NHS2013a,NHS2013b}. Although these standards establish a comprehensive clinical-safety lifecycle, their requirements focus on process transparency and documentation rather than interpretability or algorithmic transparency, which remain undefined for AI systems.

The ECSF addresses this gap without creating new artefacts. Instead, it defines acceptable forms of evidence derived from explainability techniques that can be integrated within existing DCB documentation. Global attribution and bias summaries can be incorporated into the Hazard Log to support risk identification; local case explanations can be included in the Clinical Safety Case to substantiate verification; and post-deployment explainability monitoring outputs can populate the Post-Market Surveillance Plan and associated reports described in Sections 7.2 of both DCB0129 and DCB0160.

This approach preserves compliance with existing DCB standards while operationalising the transparency and human-oversight principles articulated in the GMLP guidance and the transparency obligations of the EU AI Act \cite{MHRA2021,EUAIACT2024}.

\begin{figure}[H]
  \centering
  \includegraphics[width=\textwidth]{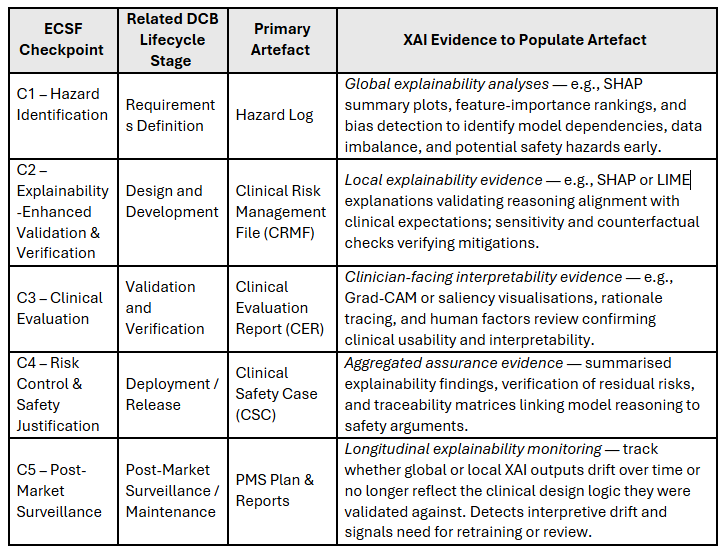} 
  \caption{Explainability-Enabled Clinical Safety Framework (ECSF) lifecycle mapping showing integration of XAI evidence across DCB0129/0160 artefacts.}
  \label{fig:ECSF_Framework}
\end{figure}

\subsection{Traditional and Deep Learning Explainability}

\textbf{Traditional machine learning.} Quantitative attribution techniques make model logic auditable and interpretable. SHAP quantifies the contribution of each feature to a given prediction, providing local interpretability; when aggregated, SHAP values provide global insights into overall feature influence and model stability, informing both the Hazard Log and Clinical Safety Case \cite{Lundberg2017}. LIME approximates model behaviour locally by generating interpretable surrogate models around individual predictions; when aggregated across samples, it approximates global feature patterns within validation cohorts \cite{Ribeiro2016}. Permutation Importance and Partial Dependence Plots extend this analysis by providing global transparency, quantifying changes in model outputs when input features are perturbed to identify dominant predictors and potential sources of bias \cite{Arrieta2020}.

\textbf{Deep learning.} Visual and gradient-based explainability methods clarify the non-linear reasoning processes of deep neural networks. Saliency mapping and Gradient-weighted Class Activation Mapping (Grad-CAM) highlight influential image regions or text tokens, providing local visual transparency that can be aggregated to reveal global focus patterns \cite{Samek2019}. Integrated Gradients quantify the cumulative contribution of each input feature along its activation path, supporting quantitative validation of model reasoning. Counterfactual and contrastive explanations illustrate how small input changes alter predictions, revealing decision boundaries and supporting human-in-the-loop usability testing \cite{Amann2020,Hildt2025,Ghassemi2021}.

\subsection{Explainability for Large Language Models and Generative AI}

Large Language Models (LLMs) require explainability approaches that address both reasoning and provenance. Evidence from the MHRA AI Airlock Pilot indicates that assurance for generative systems should combine global model transparency with local output rationale to ensure traceable and auditable behaviour across the product lifecycle \cite{Bajwa2021}.

\textbf{Global LLM explainability.} System-level traceability can be achieved through structured documentation and lifecycle transparency. Model-card reporting summarises the model’s intended use, limitations, and performance characteristics, providing a foundation for accountability and auditability \cite{Mitchell2019}. Training-data lineage documentation ensures provenance and representativeness of datasets, aligning with transparency and oversight principles in the GMLP guidance and the EU AI Act \cite{MHRA2021,EUAIACT2024}. Logging of fine-tuning parameters and model versioning during updates supports reproducibility and traceability throughout iterative development cycles \cite{Asgari2025}. Finally, specification of guardrail logic within safety documentation maintains human oversight and behavioural control, consistent with NHS AI Assurance and T.E.S.T. frameworks \cite{NHSAIAssurance2023,NHSTEST2023}.

\textbf{Local LLM explainability.} Case-level interpretability focuses on making model reasoning observable and verifiable. Rationale tracing links each output to its source context, providing a transparent evidence trail for clinical accountability \cite{MHRAAirlock2025,Asgari2025,NHSAIAssurance2023}. Token-level attention visualisation identifies which input words or data features most influenced the generated output, revealing the model’s focus and potential biases \cite{Vaswani2017,Wiegreffe2019,Hildt2025,Asgari2025}. Chain-of-thought summarisation captures intermediate reasoning steps within the generative process, enhancing transparency and supporting clinician understanding of how the model arrived at specific conclusions \cite{Wei2022,Hildt2025,Asgari2025}.

Together, these approaches enable systematic assessment of hallucination risk, contextual accuracy, and reasoning stability across the LLM lifecycle, ensuring that generative models remain clinically reliable, transparent, and accountable.

\begin{figure}[H]
\centering
\includegraphics[width=\linewidth]{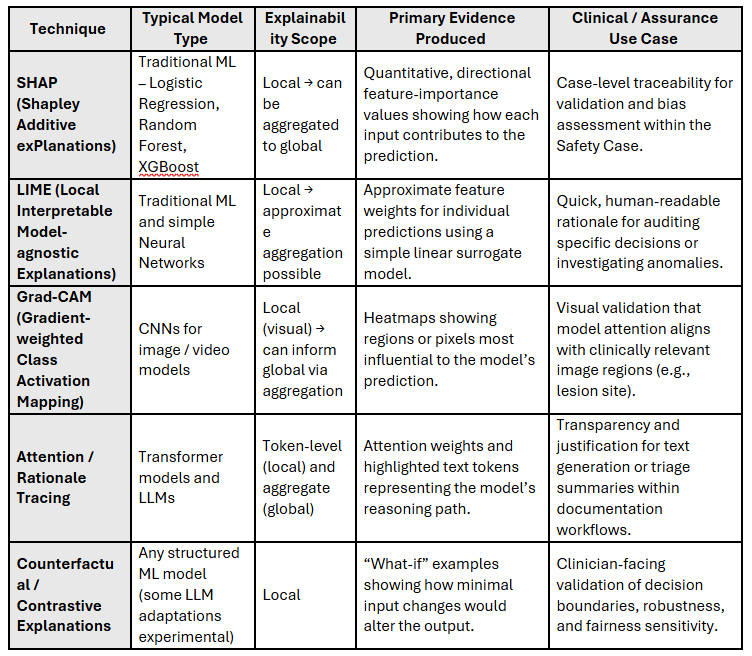}
\caption{Explainability Techniques by Model Type and Evidence Role.
This figure summarises major explainability methods (SHAP, LIME, Grad-CAM, attention/rationale tracing, and counterfactual explanations) and their evidentiary relevance across model classes.}
\label{fig:techniques}
\end{figure}

\subsection{Integration into Assurance Activities and Artefacts}

Within the ECSF, explainability evidence is systematically integrated into the five checkpoints of the DCB lifecycle. Each checkpoint aligns with existing assurance artefacts, such as the Hazard Log, Clinical Safety Case, Clinical Evaluation Report, and Post-Market Surveillance documentation, ensuring that transparency, interpretability, and oversight are evidenced throughout the system’s lifecycle. This integration preserves DCB conformance while extending the assurance process to address the probabilistic behaviour of AI systems.

\textbf{Hazard Identification – Model Transparency and Bias Detection (Global).}  
At the earliest stage, explainability supports proactive risk discovery. Global techniques such as aggregated SHAP, Permutation Importance, and Partial Dependence identify influential features, latent bias, and systemic dependencies that may lead to unsafe or inequitable outputs \cite{Lundberg2017}. For large-language and deep-learning models, equivalent system-level transparency is achieved through model cards, training-data documentation, and architecture summaries. These analyses populate the Hazard Log and form early evidence entries in the Clinical Safety Case, supporting traceability of model logic to identified risks and demonstrating compliance with DCB0129 Section 3.3 on hazard identification and evidence of risk control.

\textbf{Explainability-Enhanced Validation – Case-Level Explainability (Local).}  
During model validation, local explainability techniques demonstrate whether predictions align with expected clinical reasoning. Methods such as LIME, SHAP, Integrated Gradients, and saliency mapping clarify how specific input variables contribute to individual predictions, providing case-level interpretability that complements quantitative validation metrics \cite{Ribeiro2016}. In LLMs, token-level attention visualisation and rationale-trace comparison provide similar evidence. These outputs substantiate the verification and validation activities required under DCB0129 Section 3.5 and are incorporated into the Clinical Safety Case as supporting validation evidence.

\textbf{Clinical Evaluation – Usability and Human Oversight.}  
Explainability also supports human-factors assurance by enabling clinicians to interrogate model outputs during usability testing. Interactive dashboards, contrastive and counterfactual examples, and rationale summaries allow end users to review AI-generated recommendations, confirming that model reasoning aligns with clinical context and professional judgement \cite{Amann2020,Hildt2025}. This stage aligns with GMLP Principle 7 on the performance of the human–AI team and Principle 9 on transparency and user information. \cite{MHRA2021}. Findings are summarised within the Clinical Evaluation Report and, where interpretive or cognitive risks are identified, recorded as updates to the Hazard Log. These activities strengthen the feedback loop between technical validation and clinical usability, reinforcing interpretability as a core dimension of safety.

\textbf{Risk Mitigation and Control – Traceability of Decision Pathways (Global and Local).}  
Following validation, explainability evidence supports traceability and control verification within the Clinical Safety Case. Global and local explainability outputs, such as feature hierarchies, saliency audits, and explanation trails document how identified risks were mitigated and how control measures influence model behaviour. For LLMs, prompt-template documentation, guardrail specifications, fine-tuning records, and model-card updates provide similar traceability within configuration and change-control documentation \cite{Mitchell2019,NHSAIAssurance2023}. These records ensure that system-level and version-level changes remain auditable, supporting ongoing compliance and accountability across software iterations.

\textbf{Post-Market Surveillance – Longitudinal Explainability Monitoring (Global and Local).}  
After deployment, explainability monitoring provides longitudinal assurance of model reliability and interpretive stability. Comparative analyses of feature-importance distributions and explanation consistency detect shifts in model reasoning or data dependency over time \cite{AlvarezMelis2018,Yeh2019,Carvalho2019,Linardatos2021}. For LLMs, rationale-consistency and hallucination-rate tracking identify interpretability drift and emerging risks to output reliability \cite{Asgari2025}. These activities align with the EU AI Act’s post-market monitoring provisions (Articles 61–62) and extend the transparency obligations of Article 13 into the operational phase. They also correspond to DCB0129 Section 7.2, which requires continuous review of clinical safety and implementation of corrective actions. The resulting outputs populate the Post-Market Surveillance Plan and associated reports, ensuring that transparency and assurance are maintained beyond initial deployment.

Continuous monitoring of explainability drift addresses the nondeterministic variability observed in neural optimisation and fine-tuning, reinforcing reproducibility and reliability \cite{Bouthillier2021}. Embedding these five checkpoints within the established DCB artefacts transforms interpretability from an analytical tool into a regulated assurance mechanism. It operationalises the transparency, accountability, and oversight principles central to modern AI governance frameworks, enabling the DCB lifecycle to evidence explainability at every assurance stage \cite{Ghassemi2021}.

\begin{center}
  \includegraphics[width=\linewidth]{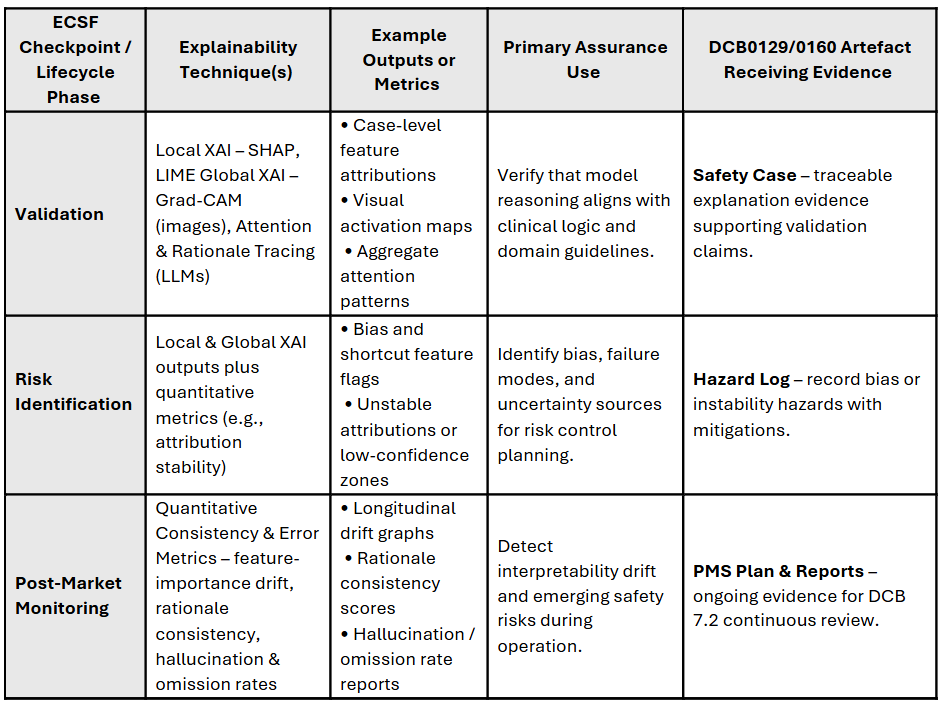}
  \captionof{figure}{This figure provides an illustrative, non-exhaustive mapping of explainability outputs (e.g., SHAP, LIME, Grad-CAM, rationale tracing, and monitoring metrics) to the corresponding stages and artefacts of the DCB0129/0160 safety lifecycle.}
  \label{fig:integration-points}
\end{center}

\subsection{Illustrative Application Scenario}

To demonstrate the practical application of the ECSF, a simulated example of an AI early-warning system for sepsis prediction was developed. The model analysed routinely collected vital signs and laboratory data to estimate sepsis risk. This example is illustrative and does not represent a live deployment; rather, it shows how explainability evidence can be systematically embedded within DCB0129 and DCB0160 artefacts across the assurance lifecycle.

\textbf{Hazard Identification.}  
An agggregate  SHAP analysis revealed disproportionate weighting of age and white blood cell count, with under-representation of lactate and respiratory rate, indicating potential model bias and an increased risk of missed atypical presentations. This issue was recorded as a new hazard in the Hazard Log and supported by a FMEA detailing the failure mode, potential clinical impact, and proposed mitigation measures. 

\textbf{Validation and Verification.}  
Following data augmentation and model retraining, updated SHAP and Partial Dependence analyses demonstrated improved feature balance, while local LIME explanations confirmed that individual predictions aligned with expected physiological reasoning. These interpretability outputs were appended to the Clinical Safety Case as validation evidence in accordance with ECSF Checkpoint~C2 (Validation and Verification). Clinicians subsequently reviewed contrastive examples through an explanation dashboard to assess contextual accuracy and confirm confidence in the model’s reasoning.

\textbf{Post-Market Surveillance.}  
Within the Post-Market Surveillance (PMS) Plan, periodic audits of explanation stability and feature-importance drift were implemented to detect changes in model reasoning over time, corresponding to ECSF Checkpoint~C5 PMS. 

Together, these activities illustrate how explainability evidence can populate existing DCB artefacts such as the Hazard Log, Clinical Safety Case, and Post Market Surveillance documentation, converting interpretability outputs into traceable and auditable clinical safety evidence consistent with NHS AI Assurance principles and  GMLP guidance. \cite{NHSAIAssurance2023,MHRA2021}.

\subsection{ECSF Alignment with Regulatory Principles and Key Outcomes}

The ECSF aligns with the primary regulatory and ethical instruments governing the safe use of artificial intelligence in the NHS. It operationalises the principles of transparency, traceability, and human oversight established in the GMLP guidelines, the EU AI Act, and the NHS AI Assurance and T.E.S.T. frameworks, while remaining consistent with the DCB0129 and DCB0160 clinical risk management standards \cite{MHRA2021,EUAIACT2024,NHSAIAssurance2023,NHSTEST2023}.

\textbf{Good Machine Learning Practice (GMLP).}  
Under the GMLP guidance, particularly Principles 7 and 9 on human–AI interaction and user transparency, AI systems are expected to generate evidence that is interpretable to intended users and auditable by regulators \cite{MHRA2021}. The ECSF addresses this by embedding explainability techniques such as SHAP, LIME, and saliency mapping within the DCB lifecycle. These methods provide global transparency by illustrating how models prioritise clinical features and local interpretability by clarifying why individual predictions occur. This evidence enables verifiable oversight, allowing Clinical Safety Officers (CSOs) to assess whether AI outputs align with clinical reasoning and to document findings within the Clinical Safety Case and Hazard Log. Beyond Principles 7 and 9, the ECSF also supports wider GMLP guidance on data quality, independence of training and test sets, and lifecycle management, ensuring that model development and maintenance are robust, bias-controlled, and traceable across the assurance process.

\textbf{EU Artificial Intelligence Act.}  
In accordance with the EU AI Act, particularly Article 13 on transparency and Articles 61–62 on post-market monitoring, high-risk AI systems must maintain comprehensive documentation and demonstrate transparency throughout their lifecycle \cite{EUAIACT2024}. The ECSF supports these obligations through the inclusion of training-data lineage, design justification, and rationale tracing to establish a continuous interpretability evidence trail. This approach ensures that retraining, fine-tuning, or model updates do not compromise transparency, reliability, or regulatory accountability.In addition to Articles 13 and 61–62, the framework reflects the broader obligations set out in Articles 9–15 of the EU AI Act, including requirements for risk management, data governance, human oversight, and accuracy.

\textbf{NHS AI Assurance and T.E.S.T. Frameworks.}  
The ECSF also reinforces the fairness, accountability, and transparency principles set out in the NHS AI Assurance and T.E.S.T. frameworks published between 2023 and 2024 \cite{NHSAIAssurance2023,NHSTEST2023}. Its five checkpoints, covering hazard identification, validation, clinical evaluation, risk control, and post-market surveillance, translate these principles into measurable assurance activities. Bias detection enhances fairness, interpretable validation strengthens transparency, and longitudinal explainability monitoring sustains accountability through continuous post-deployment review.This alignment also extends to the Security and Traceability domains of the NHS AI Assurance T.E.S.T. framework, embedding explainability outputs within controlled, auditable governance pathways.

\textbf{Integration within DCB Standards.}  
By embedding explainability outputs directly within existing DCB artefacts, the ECSF extends the scope of DCB0129 and DCB0160 without altering their compliance pathways. Global and local, specific evidence are incorporated into the formal documentation, linking model reasoning to clinical risk management in a reproducible and auditable manner \cite{NHS2013a,NHS2013b}.

\begin{figure}[H]
\centering
\includegraphics[width=\linewidth]{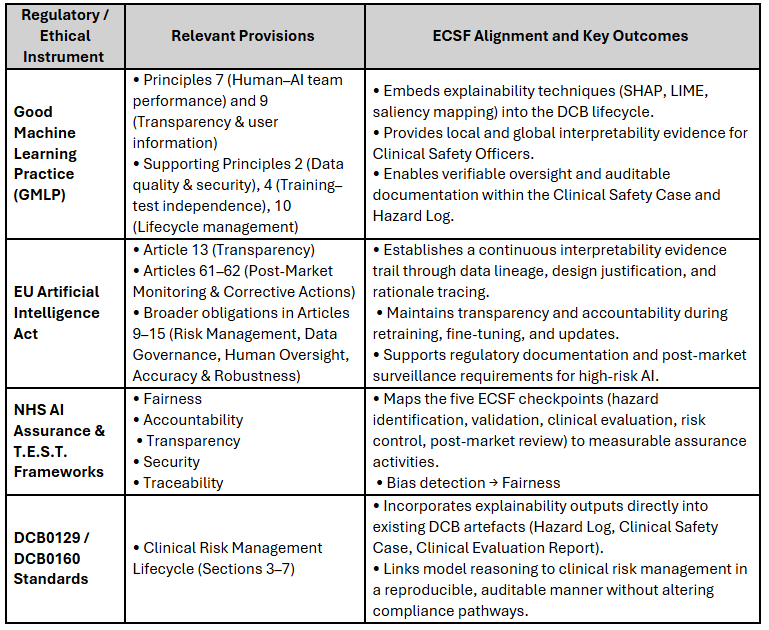}
\caption{Governance Alignment Matrix summarising conceptual correspondence between ECSF checkpoints, 
Good Machine Learning Practice (GMLP) principles, EU AI Act provisions (Articles 9–15, 61–62), 
and NHS AI Assurance Framework domains (T.E.S.T.). 
The table illustrates how ECSF checkpoints operationalise transparency, risk management, 
and post-market assurance across multiple governance frameworks.}
\label{fig:governance-alignment}
\end{figure}

\textbf{Key Outcomes.}  
This regulatory alignment delivers tangible outcomes: it enhances traceability by enabling reconstruction of model reasoning, strengthens safety evidence chains through interpretable outputs, improves clinician oversight through interactive rationale tracing, and supports continuous assurance through monitoring of explanation stability and drift. Collectively, these outcomes improve regulatory and ethical readiness by ensuring that audits rely on transparent, standardised interpretability evidence rather than retrospective justification \cite{Bajwa2021,Chustecki2024,Amann2020}.

By embedding explainability across the assurance lifecycle, the ECSF reframes interpretability from a technical aspiration into a regulated safety function that bridges model transparency with clinical accountability in AI-enabled healthcare.

\section{Discussion}

\subsection{Reframing Explainability as a Safety Enabler}

The ECSF provides a pragmatic mechanism for translating abstract interpretability research into operational clinical-safety practice. Traditional NHS assurance processes, developed for deterministic software, assume consistent performance under fixed parameters. However, the adaptive and probabilistic behaviour of AI systems challenges these assumptions. Even under controlled conditions, model outputs can vary across runs or datasets due to stochastic optimisation processes and sampling effects \cite{Bouthillier2021,Atil2025}. This inherent variability complicates verification and reproducibility, requiring new assurance mechanisms that make uncertainty observable and manageable rather than eliminated.

Explainability provides this mechanism by transforming AI from an opaque computational process into an auditable system. Rather than seeking to remove uncertainty, the ECSF manages it through structured visibility. Explainability checkpoints embedded within the DCB0129 and DCB0160 lifecycle document, interpret, and contextualise model behaviour, converting opacity into evidence. This allows CSOs to trace model reasoning, assess alignment with clinical intent, and record findings as formal assurance evidence \cite{Holzinger2019a,Tonekaboni2019} In doing so, the ECSF reframes explainability as an active safety control, consistent with NHS AI Assurance principles on transparency and human oversight \cite{NHSAIAssurance2023,MHRA2021}.

\subsection{Strengthening Traceability, Validation, and Accountability}

Ensuring traceability, the ability to reconstruct how a model reaches specific outputs and which features influence its decisions remains one of the greatest challenges in AI assurance. Explainability methods such as SHAP, LIME, and attention visualisation enable this traceability at both the local  and global scales \cite{Ribeiro2016,Lundberg2017,Samek2019}. Although the faithfulness and completeness of explanations remain important evaluation concerns, the ECSF treats interpretability evidence as corroborative rather than conclusive, complementing performance, bias, and safety testing \cite{Carvalho2019,Linardatos2021}.

Within the ECSF, explainability outputs form core safety evidence rather than supplementary material. During validation, interpretability demonstrates whether model reasoning aligns with expected clinical logic. During post-market surveillance, shifts in explanations or rationale indicate emerging bias or performance drift, including hallucination-related issues in large language models \cite{Asgari2025}. During incident investigation, visual and quantitative explanations help identify whether failures arise from data artefacts, bias, or model instability. 

By integrating these interpretability outputs into DCB artefacts such as the Hazard Log and Clinical Safety Case, the ECSF embeds traceability and accountability across the lifecycle. This structure reinforces the oversight principles central to the GMLP, the EU AI Act, and the NHS AI Assurance Framework \cite{MHRA2021,EUAIACT2024,NHSAIAssurance2023}.

\subsection{Addressing Ethical and Human Factors}

Explainability functions not only as a technical safeguard but also as an ethical and human-centred requirement. Clinicians are more likely to trust and adopt AI systems when they can interpret and challenge a model’s reasoning. Conversely, opaque models risk hesitation, misuse, or overreliance, particularly when outputs conflict with professional judgement. Contextual and actionable explanations improve confidence, support shared decision-making, and mitigate automation bias \cite{Tonekaboni2019,Hildt2025}.

From an ethical perspective, explainability reinforces the principles of autonomy, non-maleficence, and accountability by enabling clinicians to understand and justify AI-assisted recommendations within professional and patient-safety norms. However, explanations must also communicate uncertainty to avoid false confidence \cite{Ghassemi2021}. When uncertainty is high, rationale summaries should explicitly disclose model confidence levels and limitations.

Explainability must remain clinically usable. Overly complex or inconsistent outputs can increase cognitive burden and erode clinician trust. The ECSF therefore promotes a tiered approach: simplified outputs for clinicians, detailed audit trails for CSOs, and technical diagnostics for developers. This layered structure aligns with the NHS AI Assurance Framework’s emphasis on clarity and proportionality, ensuring that interpretability enhances rather than hinders safe clinical adoption \cite{NHSAIAssurance2023}.

\subsection{Regulatory Convergence and Practical Alignment}

The ECSF addresses the fragmentation of AI safety and assurance guidance across regulatory bodies. While DCB0129 and DCB0160 govern NHS clinical safety, complementary frameworks such as GMLP, the EU AI Act, and the NHS AI Assurance Framework articulate overlapping but distinct principles of transparency, fairness, and accountability. The ECSF acts as a bridge between these instruments, translating high-level principles into practical, auditable mechanisms embedded within clinical governance documentation \cite{Bajwa2021,Chustecki2024}.

For example, the transparency principle of the GMLP is operationalised through ECSF explainability checkpoints, while the EU AI Act’s documentation and oversight obligations are fulfilled through traceable interpretability evidence. Similarly, the NHS AI Assurance Framework’s fairness and accountability requirements are reinforced through bias detection and longitudinal explainability monitoring. By embedding explainability evidence within DCB artefacts such as the Clinical Safety Case, Hazard Log, and Post-Market Surveillance Plan, the ECSF ensures that these frameworks function as integrated components of a single safety ecosystem. This convergence reduces duplication, enhances regulatory coherence, and enables local governance teams to meet both national and international assurance expectations simultaneously.

\subsection{Technical and Empirical Limitations}

Although the ECSF provides a coherent conceptual model, its practical effectiveness has yet to be validated empirically. Real-world implementation will require iterative evaluation and multidisciplinary collaboration. Several limitations must be acknowledged.

\textbf{Stability and Reproducibility of Explainability Methods.}  
Techniques such as SHAP and LIME can yield variable results depending on sampling strategy or model configuration \cite{Bouthillier2021,Summers2021}. The ECSF therefore recommends reporting explanation-stability metrics and maintaining version-controlled evidence to ensure reproducibility.

\textbf{Nondeterminism and Model Variability.}  
Neural and large language models may generate different outputs across runs or software versions \cite{Atil2025}. The ECSF mitigates this through continuous explainability logging, treating interpretability as a dynamic evidence trail rather than a static output.

\textbf{Model and Data Drift.}  
Real-world data shifts can affect both model performance and the consistency of its explanations. The ECSF’s post-market checkpoint highlights the need to establish drift thresholds and re-evaluation criteria within the lifecycle to maintain model validity and clinical safety. \cite{NHSAIAssurance2023}.

\textbf{Manipulation and Prompt Instability in LLMs.}  
Prompt or guardrail changes can alter rationales without changing apparent accuracy. The ECSF mandates prompt and guardrail versioning, alongside rationale-consistency audits, to prevent inadvertent or malicious manipulation \cite{MHRAAirlock2025}.

\textbf{Clinician Usability and Cognitive Load.}  
Overly detailed explanations risk overwhelming clinical users. Implementation should prioritise interpretive clarity through adaptive dashboards and context-aware visualisations \cite{Hildt2025}.

\textbf{Ethical Ambiguity.}  
Explainability alone does not ensure fairness or safety, transparent models can still embed bias or harm. Therefore, explainability must operate alongside bias audits, validation testing, and accountability mechanisms \cite{Chustecki2024,Amann2020}.

\textbf{Empirical Validation.}  
The ECSF has yet to be tested in live NHS environments. Pilot evaluations within the NHS AI Airlock could assess feasibility, usability, and regulatory impact in practice \cite{MHRAAirlock2025}.

\subsection{Pathways for Implementation}

Successful implementation of the ECSF will require coordinated action across three operational domains.

\textbf{Technical Integration.}  
Integrate explainability tools such as SHAP, LIME, and attention visualisation directly into development and assurance pipelines. All explainability artefacts such as plots, metrics, or rationale traces, should be version-controlled and linked to model or data hashes within Clinical Safety Case revisions to ensure traceability and reproducibility.

\textbf{Governance Alignment.}  
Embed explainability checkpoints within DCB0129 and DCB0160 workflows, ensuring they are explicitly referenced in the Hazard Log, Clinical Safety Case, and Post-Market Surveillance Plan. Drift thresholds and change-control criteria should be defined in accordance with NHS AI Assurance and  GMLP expectations \cite{NHSAIAssurance2023,MHRA2021}. These measures ensure that explainability evidence remains aligned with regulatory governance and lifecycle management.

\textbf{Training and Organisational Culture.}  
Equip CSOs, developers, and governance teams with the skills required to interpret and apply explainability outputs in safety management. Fostering a culture of transparency will normalise interpretability as a routine component of clinical safety rather than a technical add-on.

Initial implementation should prioritise lower-risk applications such as clinical documentation support or triage tools, before progressing to diagnostic or treatment-planning systems. This phased approach enables refinement of explainability processes and assurance mechanisms prior to their application in higher-stakes clinical domains.

\subsection{Implications for Future Regulation and Research}

The ECSF contributes to the emerging international consensus that explainability should be recognised as a regulatory and clinical-safety enabler rather than a purely technical goal. By embedding explainability within the DCB lifecycle, it supports post-market monitoring, human oversight, and traceability requirements articulated in the EU AI Act and anticipated in future UK Medical Device Regulations \cite{EUAIACT2024}.

Future research should evaluate the practical usability and interpretive accuracy of explainability outputs within clinical safety workflows, as well as the quantitative stability and reproducibility of explainability methods across datasets and model architectures. Further studies are needed to assess the impact of explainability on clinician confidence and decision quality, and to explore opportunities for standardising interpretability evidence across suppliers to enable consistent NHS assurance audits. Comparative benchmarking of explanation-stability and hallucination-rate metrics across vendors and datasets will also be essential \cite{Asgari2025}.

\section{Conclusion}

This paper proposes the ECSF, which embeds explainability as routine evidence within the DCB0129 and DCB0160 lifecycle. By aligning established interpretability techniques with existing NHS clinical safety artefacts, the framework converts model reasoning into auditable assurance evidence that supports hazard identification, validation and verification, clinical evaluation, risk control, and post-market surveillance. In doing so, it bridges deterministic safety processes with the probabilistic behaviour of AI systems and operationalises the principles of transparency, traceability, and human oversight articulated in current UK and EU guidance \cite{MHRA2021,EUAIACT2024,NHSAIAssurance2023}.

The framework’s practical contribution lies in specifying where and how global and local model-specific explanations populate the Hazard Log, Clinical Safety Case, Clinical Evaluation Report, and Post-Market Surveillance documentation. This strengthens traceability of decision pathways, improves the evidentiary basis of safety arguments, and supports continuous monitoring over time. It also provides a structured mechanism for organisations to meet converging expectations across NHS AI Assurance,  GMLP, and the EU AI Act while maintaining full compatibility with current DCB workflows \cite{NHS2013a,NHS2013b,MHRA2021,EUAIACT2024}.

Two limitations remain. First, the ECSF is conceptual and requires empirical evaluation in live  environments to assess feasibility, usability, and its effect on clinical safety outcomes. Second, explanation methods may be unstable or incomplete; therefore, explainability evidence should complement rather than replace performance, bias, and safety testing.

Future work should prioritise pilot implementations, to generate comparative evidence on explanation stability, clinician interpretability, and auditability across suppliers. Over time, such evaluations could inform the development of standardised templates for explainability evidence within DCB artefacts, supporting a unified and proportionate approach to AI assurance in the NHS \cite{MHRAAirlock2025,NHSAIAssurance2023}.The next phase of this work will focus on implementing a prototype machine-learning model to empirically populate the ECSF artefacts, testing how explainability outputs (e.g., SHAP, LIME, rationale tracing) can serve as verifiable clinical-safety evidence within the DCB0129/0160 framework.

\section*{Ethics approval}
Ethical approval was not required as this study did not involve human participants or identifiable data.

\section*{Funding}
The author received no financial support for the research, authorship, and/or publication of this article.

\section*{Competing interests}
The author declares no conflicts of interest.

\bibliographystyle{vancouver}
\bibliography{ecsf_references}
\end{document}